\documentclass[a4paper,12pt]{article}
\usepackage{graphicx}
\usepackage{mathtools}
\usepackage{amssymb}
\usepackage{amsfonts}
\usepackage[footnotesize]{caption}
\usepackage[font=scriptsize]{subcaption}
\usepackage{color}
\usepackage{braket}
\usepackage{cite}
\usepackage{hyperref}
\usepackage{url}
\usepackage{multirow}
\usepackage{relsize}
\usepackage{fullpage}
\usepackage{makecell}
\usepackage{blkarray}
\usepackage{fullpage}

\newcommand{\newc}{\newcommand}
\newc{\be}{\begin{equation}}
\newc{\ee}{\end{equation}}
\newc{\bea}{\begin{eqnarray}}
\newc{\eea}{\end{eqnarray}}
\newc{\simlt}{~\mbox{\smaller\(\lesssim\)}~}
\newc{\simgt}{~\mbox{\smaller\(\gtrsim\)}~}

\newcommand{\pmatr}[1]{\begin{pmatrix} #1 \end{pmatrix}}

\begin{document}

\begin{titlepage}
\begin{center}
{\bf\Large
\boldmath{
Flavourful $Z'$ models for $R_{K^{(*)}}$
}
} \\[12mm]
Stephen~F.~King$^{\star}$%
\footnote{E-mail: \texttt{king@soton.ac.uk}}
\\[-2mm]
\end{center}
\vspace*{0.50cm}
\centerline{$^{\star}$ \it
School of Physics and Astronomy, University of Southampton,}
\centerline{\it
SO17 1BJ Southampton, United Kingdom }
\vspace*{1.20cm}

\begin{abstract}
{\noindent
We show how any flavour conserving $Z'$ model
can be made flavour violating and non-universal 
by introducing mass mixing of quarks and leptons with a fourth family of vector-like fermions with non-universal $Z'$ couplings.
After developing a general formalism, we
focus on two concrete examples, namely a fermiophobic model, and an 
$SO(10)$ GUT model, and show how they can account for 
the anomalous $B$ decay ratios
$R_K$ and $R_{K^*}$. 
A similar analysis could be performed for $B-L$ models, $E_6$ models,
composite models, and so on.
}
\end{abstract}
\end{titlepage}

\section{Introduction}

One of the simplest extensions of the Standard Model (SM) is to introduce an additional gauged $U(1)'$,
which could emerge 
as a remnant of larger gauge group embeddings of the SM gauge group, with rank larger than 4.
Such larger gauge groups include the left-right symmetric model, Pati-Salam, $SO(10)$,
$E_6$. An extra gauged $U(1)'$ is common in string inspired models, where it is difficult to break the rank of the gauge group,
or from alternative dynamical schemes such as 
composite models.
For a review of $Z'$ models and an extensive list of references see e.g. \cite{Langacker:2008yv}.

Most of the existing $Z'$ models have universal couplings to the three families of quarks and leptons.
The reason for this is both theoretical and phenomenological. Firstly many theoretical models naturally 
predict universal $Z'$ couplings. Secondly, 
from a phenomenological point of view, having universal couplings avoids
dangerous favour changing neutral currents (FCNCs) mediated by tree-level $Z'$ exchange.
The most sensitive processes 
involve the first two families, such as $K_0-\bar K_0$
mixing, $\mu - e$ conversion in muonic atoms, and so on, leading to stingent bounds on the $Z'$ mass and couplings
 \cite{Langacker:2008yv}.

Recently, the phenomenological motivation for considering non-universal $Z'$ models has increased due to 
mounting evidence for semi-leptonic $B$ decays which violate $\mu - e$ universality at rates which exceed those predicted by the SM
\cite{Descotes-Genon:2013wba}.
In particular, the LHCb Collaboration and other experiments have reported a number of anomalies in $B\rightarrow K^{(*)}l^+l^-$
decays such as the $R_K$ \cite{Aaij:2014ora} and $R_{K^*}$ \cite{Bifani} ratios of $\mu^+ \mu^-$ to $e^+ e^-$ final states, 
which are observed to be about $70\%$ of their expected values with a $4\sigma$ deviation from the SM,
and the $P'_5$ angular variable,
not to mention the $B\rightarrow \phi \mu^+ \mu^-$ mass distribution in $m_{\mu^+ \mu^-}$. 

Following the recent measurement of $R_{K^*}$ \cite{Bifani}, a number of phenomenological analyses of these data, 
see e.g. \cite{Hiller:2017bzc},
favour a operator of the left-handed (L) form \cite{Glashow:2014iga}, 
in the conventions of \cite{DAmico:2017mtc},
\be
V_{tb}V_{ts}^*\frac{\alpha_{em}}{4\pi v^2}
\left(C^{\rm SM}_{b_L\mu_L} + C^{\rm BSM}_{b_L\mu_L}\right)  \bar b_L\gamma^{\mu} s_L \, \bar \mu_L \gamma_{\mu} \mu_L
\label{1}
\ee
where the SM operator arises from penguin diagrams and has a coefficient of $C^{\rm SM}_{b_L\mu_L}=8.64$,
while the beyond the SM (BSM) operator has a coefficient of $C^{\rm BSM}_{b_L\mu_L}\approx -1.3$.
The analogous right-handed (R) operators must be significantly smaller \cite{DAmico:2017mtc}.
The SM constants $V_{ts}=0.040\pm 0.001$ (predominantly real) and the Higgs vacuum expectation value (VEV) $v=174$ GeV, set
the scale of Eq.\ref{1},
\be
V_{tb}V_{ts}^*\frac{\alpha_{em}}{4\pi v^2}\approx \frac{1}{(36 \ {\rm TeV})^2} .
\ee
This suggests a new physics operator of the form,
\be
G^{\rm BSM}_{b_L\mu_L} \;   \bar b_L\gamma^{\mu} s_L \, \bar \mu_L \gamma_{\mu} \mu_L \approx -\frac{1}{(33 \ {\rm TeV})^2} \; \bar b_L\gamma^{\mu} s_L \, \bar \mu_L \gamma_{\mu} \mu_L .
\label{G}
\ee
In a flavourful $Z'$ model, the new physics operator in Eq.\ref{G} will arise from tree-level $Z'$ exchange,
where the $Z'$ must dominantly couple to $\mu_L \mu_L$ over $\mu_R \mu_R$, $e_L e_L$, $e_R e_R$,
and must also have the quark flavour changing coupling $b_L s_L$ which must dominate over $b_R s_R$.
The coefficient of the tree-level $Z'$ exchange operator will be typically of the form,
\be
G^{\rm BSM}_{b_L\mu_L} \; = g^{Z'}_{b_L} g^{Z'}_{\mu_L} \left(\frac{g'^2}{{M_{Z'}}^2}\right)  \approx -\frac{1}{(33 \ {\rm TeV})^2}
\label{33}
\ee
where the Feynman rule for the $Z' \bar b_L\gamma^{\mu} s_L$ coupling is $-i\gamma^{\mu} g^{Z'}_{b_L}g'$ and the 
$Z' \bar \mu_L \gamma^{\mu} \mu_L$ coupling is $-i \gamma^{\mu} g^{Z'}_{\mu_L}g'$,
where $g'$ is the $Z'$ gauge coupling and $M_Z'$ is the mass of the $Z'$.
The required value of $M_Z'$ will typically be much smaller than $33 \ {\rm TeV}$ due to the model dependent coupling
factors $g^{Z'}_{b_L}$ and $g^{Z'}_{\mu_L}$ which are anticipated to be quite small in realistic models.
This means that the $Z'$ in these models may be within reach of the LHC.

Motivated by the above considerations, 
there has been a large and growing body of literature which is concerned with flavour dependent $Z'$ models (see e.g.~\cite{large}). 
Recent works on flavoured $Z'$ approaches following the $R_{K^*}$ measurement include those in~\cite{recent}.
One of the key challenges faced by these models is the requirement that they be anomaly free.
This has motivated the phenomenological analysis of $Z'$ models based on gauged $L_{\mu}-L_{\tau}$,
possibly combined with vector-like quarks
\cite{Bobeth:2016llm}. Without a $Z'$, vector-like quarks directly mixing with ordinary quarks via
the Higgs Yukawa couplings can lead to FCNCs \cite{delAguila:2000rc}. However, vector-like quarks with a gauged $U(1)'$
typically forbids the Higgs coupling of vector-like quarks to ordinary quarks, but allows new possibilities \cite{Bobeth:2016llm}.
For example, a simple idea is to have a dark $U(1)_X$ under which the SM quarks and leptons are neutral,
but which is felt by vector-like fermions with the SM quantum numbers of the doublets
$Q_L$ and $L_L$, leading to a dark matter candidate and flavour-changing $Z'$ operators after the vector-like fermion mass terms
mix with SM fermions
 \cite{Sierra:2015fma}. However adding such matter spoils the prospects for gauge coupling unification unless the vector-like matter comes in complete representations of $SU(5)$. The first example of mixing with vector-like fermions which preserves gauge unification
 and leads to flavour-changing $Z'$ interactions was proposed some time ago
 by Langacker and London \cite{Langacker:1988ur}.
 
In this paper, motivated by the $R_K$ and $R_{K^*}$ anomalies, we show how any flavour conserving $Z'$ model
can be made flavour violating and non-universal 
by the mass mixing of quarks and leptons with a fourth family of vector-like fermions with non-universal $Z'$ couplings.
Unlike the original vector-like fermion models \cite{delAguila:2000rc}, having non-universal $U(1)'$ charges of the fourth vector-like family 
forbids mixing via the usual Higgs Yukawa couplings. Instead, new singlet scalars with appropriate $U(1)'$ charges
are added to generate mass mixing of quarks and leptons with the vector-like family.
Since we include a complete vector-like family, the mixing will include the doublets
$Q_L$ and $L_L$, leading to the left-handed new physics operators required for $R_K$ and $R_{K^*}$.
Since we consider a complete fourth vector-like family, unification is maintained.
We develop a quite general formalism, which can be applied to any $Z'$ model in the literature, including $B-L$ models, $E_6$ models,
composite models, and so on.
To illustrate the mechanism we consider two concrete examples, namely a fermiophobic model, and an 
$SO(10)$ Grand Unified Theory (GUT), and show how they can account for 
the anomalous $B$ decay ratios
$R_K$ and $R_{K^*}$. 
 
The layout of the remainder of the paper is as follows.
In section~\ref{general} we consider the general 
class of models consisting of the usual 
three chiral families of left-handed quarks and leptons with one (or two)
Higgs doublet(s) $H_{(u,d)}$,
plus a fourth 
vector-like family of fermions, which has non-universal charges under a gauged $U(1)'$.
We write down the Lagrangian for such a general class of models in the charge basis and the heavy mass
basis, after diagonalisation of the heavy masses.
In section~\ref{examples}, to illustrate the mechanism and how it may be applied in practice,
we consider two concrete examples of well known $Z'$ models which can be made flavourful
via mixing with a non-universal fourth vector-like family, namely a
namely a fermiophobic model, and an 
$SO(10)$ GUT model, and show how they can account for 
the anomalous $B$ decay ratios
$R_K$ and $R_{K^*}$. 
Section~\ref{conclusion} concludes the paper.

\section{A class of $Z'$ models with a vector-like family}
\label{general}

\begin{table}
\centering
\footnotesize
\captionsetup{width=0.9\textwidth}
\begin{tabular}{| c c c c c  |}
\hline
\multirow{2}{*}{\rule{0pt}{4ex}Field}	& \multicolumn{4}{c |}{Representation/charge} \\
\cline{2-5}
\rule{0pt}{3ex}			& $SU(3)_c$ & $SU(2)_L$ & $U(1)_Y$ &$U(1)'$ \\ [0.75ex]
\hline \hline
\rule{0pt}{3ex}%
$Q_{Li}$ 		 & ${\bf 3}$ & ${\bf 2}$ & $1/6$ & $q_{Q_i}$\\
$u_{Ri}$ 		 & ${\bf 3}$ & ${\bf 1}$ & $2/3$ & $q_{u_i}$\\
$d_{Ri}$ 		 & ${\bf 3}$ & ${\bf 1}$ & $-1/3$ & $q_{d_i}$\\
$L_{Li}$ 		 & ${\bf 1}$ & ${\bf 2}$ & $-1/2$ & $q_{L_i}$\\
$e_{Ri}$ 		 & ${\bf 1}$ & ${\bf 1}$ & $-1$ & $q_{e_i}$\\
$\nu_{Ri}$         & ${\bf 1}$ & ${\bf 1}$ & $0$ & $q_{\nu_i}$\\
\hline
\hline
\rule{0pt}{3ex}%
$H_u$ & ${\bf 1}$ & ${\bf 2}$ & $-1/2$ &$q_{H_u}$ \\
$H_d$ & ${\bf 1}$ & ${\bf 2}$ & $1/2$ &$q_{H_d}$ \\
\hline
\hline
\rule{0pt}{3ex}%
$Q_{L4}$,$\tilde{Q}_{R4}$   		 & ${\bf 3}$ & ${\bf 2}$ & $1/6$ & $q_{Q_4}$\\
$u_{R4}$,$\tilde{u}_{L4}$  		 & ${\bf 3}$ & ${\bf 1}$ & $2/3$ & $q_{u_4}$\\
$d_{R4}$,$\tilde{d}_{L4}$  		 & ${\bf 3}$ & ${\bf 1}$ & $-1/3$ & $q_{d_4}$\\
$L_{L4}$,$\tilde{L}_{R4}$  		 & ${\bf 1}$ & ${\bf 2}$ & $-1/2$ & $q_{L_4}$\\
$e_{R4}$,$\tilde{e}_{L4}$  		 & ${\bf 1}$ & ${\bf 1}$ & $-1$ & $q_{e_4}$\\
$\nu_{R4}$,$\tilde{\nu}_{L4}$  		 & ${\bf 1}$ & ${\bf 1}$ & $0$ & $q_{\nu_4}$\\
\hline
\hline
\rule{0pt}{3ex}%
$\phi_{Q,u,d,L,e}$ & ${\bf 1}$ & ${\bf 1}$ & $0$ &$q_{\phi_{Q,u,d,L,e}}$ \\
\hline
\end{tabular}
\caption{The most general model we consider consists of the usual 
three chiral families of left-handed (L) and right-handed (R) quarks and leptons $\psi_i$ ($i=1,2,3$) and one (or two)
Higgs doublet(s) $H_{(u,d)}$,
plus a fourth 
vector-like family of fermions $\psi_4$, $\tilde{\psi}_4$.
There may be other exotics in addition to those shown in order to cancel anomalies,
or the three chiral families may cancel anomalies by themselves without extra exotics.
In any case, the vector-like fermion family are always anomaly free by themselves.
The $U(1)'$ is broken by the VEVs of new Higgs singlets $\phi_{\psi}$ 
with charges $|q_{\phi_\psi}| =|q_{\psi_i} - q_{\psi_4} |$ to yield a massive $Z'$.
}
\label{tab:funfields}
\end{table}
In this section we analyse the general 
class of models defined in Table~\ref{tab:funfields} consisting of the usual 
three chiral families of left-handed (L) and right-handed (R) quarks and leptons $\psi_i$ ($i=1,2,3$) and one (or two)
Higgs doublet(s) $H_{(u,d)}$,
plus a fourth 
vector-like family of fermions $\psi_4$, $\tilde{\psi}_4$.
The gauged $U(1)'$ charges $q_{\psi_i}$ are universal up to the fourth
family (i.e. $q_{\psi_1} =q_{\psi_2} =q_{\psi_3} \neq q_{\psi_4} $), although in general they need not be.
The three chiral families must be anomaly free, since the vector-like family is anomaly free.
The $U(1)'$ is broken by the VEVs of new Higgs singlets $\phi_{\psi}$ 
with charges $|q_{\phi_\psi}| =|q_{\psi_i} - q_{\psi_4} |$ to yield a massive $Z'$.

The layout of this rather lengthy section is as follows.
In the first subsection we present the Lagrangian of the general class of models in the charge basis.
In the second subsection we show how the heavy masses may be diagonalised.
In the third subsection we present the Lagrangian of the general class of models in the heavy mass basis.

\subsection{The Lagrangian in the charge basis}

In this subsection we present the Lagrangian of the general class of models in the charge basis.
Including the fourth family, along with the usual three chiral families,
the gauge part of the Lagrangian involving fermions is given by,
\bea
{\cal L}^{gauge}&=&i \overline{Q}_{L\alpha}\left(\partial_{\mu} - i g_3G^A_{\mu}\frac{\lambda^A}{2}
- i g_2W^a_{\mu}\frac{\sigma^a}{2}-\frac{1}{6}ig_1B_{\mu}
-q_{Q\alpha}ig'B'_{\mu}\right)\gamma^{\mu}{Q}_{L\alpha} \nonumber \\
&+&
i \overline{u}_{R\alpha}\left(\partial_{\mu} - i g_3G^A_{\mu}\frac{\lambda^A}{2}
 -\frac{2}{3}ig_1B_{\mu}
-q_{u\alpha}ig'B'_{\mu}\right)\gamma^{\mu}{u}_{R\alpha} \nonumber \\
&+&
i \overline{d}_{R\alpha}\left(\partial_{\mu} - i g_3G^A_{\mu}\frac{\lambda^A}{2}
 +\frac{1}{3}ig_1B_{\mu}
-q_{d\alpha}ig'B'_{\mu}\right)\gamma^{\mu}{d}_{R\alpha} \nonumber \\
&+&
i \overline{L}_{L\alpha}\left(\partial_{\mu} - i g_2W^a_{\mu}\frac{\sigma^a}{2}+\frac{1}{2}ig_1B_{\mu}
-q_{L\alpha}ig'B'_{\mu}\right)\gamma^{\mu}{L}_{L\alpha} \nonumber \\
&+&
i \overline{e}_{R\alpha}\left(\partial_{\mu} +ig_1B_{\mu}
-q_{e\alpha}ig'B'_{\mu}\right)\gamma^{\mu}{e}_{R\alpha}  \nonumber \\
&+&
i \overline{\nu}_{R\alpha}\left(\partial_{\mu} -q_{\nu\alpha}ig'B'_{\mu}\right)\gamma^{\mu}{\nu}_{R\alpha} 
\label{gauge}
\eea
where $\alpha =1,\ldots ,4$ labels the four families of the same chirality, 
$G^A_{\mu}$ are $SU(3)_c$ gauge fields (the octet of gluons $A =1,\ldots ,8$), $W^a_{\mu}$ are $SU(2)_L$ gauge fields
($a =1,\ldots ,3$), $B_{\mu}$
are the $U(1)_Y$ gauge fields and $B'_{\mu}$
are the $U(1)'$ gauge fields, with the three usual gauge couplings $g_i$, as well as the $U(1)'$ gauge coupling $g'$.
We denote the Pauli matrices as $\sigma^a$ and the Gell-Mann matrices as $\lambda^A$.

In addition there is a similar gauge Lagrangian involving the fourth family of the opposite chirality
$\tilde{\psi}_4$, obtained by the replacements,
${Q}_{L\alpha}\rightarrow \tilde{Q}_{R4}$, ${u}_{R\alpha}\rightarrow \tilde{u}_{L4}$, ${d}_{R\alpha}\rightarrow \tilde{d}_{L4}$,
${L}_{L\alpha}\rightarrow \tilde{L}_{R4}$, ${e}_{R\alpha}\rightarrow \tilde{e}_{L4}$,
${\nu}_{R\alpha}\rightarrow \tilde{\nu}_{R4}$.

The right-handed neutrinos are special, since the Standard Model gauge group allows
large Majorana masses, although these may be forbidden by $U(1)'$.
Henceforth, for simplicity, we shall ignore the right-handed neutrinos, and the associated vector-like fourth family,
which is equivalent to ignoring neutrino mass.

We assume that the $U(1)'$ charges allow for 
Yukawa couplings of the first three chiral families $\psi_i$, but not the fourth vector-like family,
\be
{\cal L}^{Yuk}=y^u_{ij}H_u\overline{Q}_{Li}u_{Rj}+y^d_{ij}H_d\overline{Q}_{Li}d_{Rj}
+y^e_{ij}H_d\overline{L}_{Li}e_{Rj} +H.c.
\label{yuk}
\ee
where $i,j=1,\ldots ,3$.

We assume that the $U(1)'$ charges allow for
the fourth opposite chirality family $\tilde{\psi}_4$ to have interactions with the first three chiral families $\psi_i$ via singlet fields $\phi$
which carry $U(1)'$ charge,
in addition to explicit masses between opposite chirality fourth family fields $\tilde{\psi}_4$ and $\psi_4$
of the same charges,
\bea
{\cal L}^{mass}&=&x^{Q}_{i}\phi_Q \overline{Q}_{Li}\tilde{Q}_{R4}
+x^{u}_{i}\phi_u \overline{\tilde{u}}_{L4} u_{Ri}
+x^{d}_{i}\phi_d \overline{\tilde{d}}_{L4} d_{Ri}
+x^{L}_{i}\phi_L \overline{L}_{Li}\tilde{L}_{R4}
+x^{e}_{i}\phi_e \overline{\tilde{e}}_{L4} e_{Ri}
\nonumber \\
&+& M^{Q}_{4}\overline{Q}_{L4}\tilde{Q}_{R4}
+M^{u}_{4}\overline{\tilde{u}}_{L4} u_{R4}
+M^{d}_{4}\overline{\tilde{d}}_{L4} d_{R4}
+M^{L}_{4} \overline{L}_{L4}\tilde{L}_{R4}
+M^{e}_{4} \overline{\tilde{e}}_{L4} e_{R4}
+H.c.
\eea
After the singlet fields $\phi$ develop vacuum expectation values (VEVs), we may define
new mass parameters $M^Q_i=x^Q_i\langle \phi_Q \rangle$, and similarly for the other mass parameters,
to give,
\be
{\cal L}^{mass}=
M^{Q}_{\alpha}\overline{Q}_{L\alpha}\tilde{Q}_{R4}
+M^{u}_{\alpha}\overline{\tilde{u}}_{L4} u_{R\alpha}
+M^{d}_{\alpha}\overline{\tilde{d}}_{L4} d_{R\alpha}
+M^{L}_{\alpha} \overline{L}_{L\alpha}\tilde{L}_{R4}
+M^{e}_{\alpha} \overline{\tilde{e}}_{L4} e_{R\alpha}
+H.c.
\ee
where $\alpha =1,\ldots ,4$.

\subsection{Diagonalising the heavy masses}
In this subsection we show how the heavy masses may be diagonalised, denoting the fields in this basis by primes.
The idea is that, after diagonalisation, only the fourth family is massive (before electroweak symmetry breaking),
\be
{\cal L}^{mass}=
\tilde{M}^{Q}_{4}\overline{Q'}_{L4}\tilde{Q}_{R4}
+\tilde{M}^{u}_{4}\overline{\tilde{u}}_{L4} u'_{R4}
+\tilde{M}^{d}_{4}\overline{\tilde{d}}_{L4} d'_{R4}
+\tilde{M}^{L}_{4} \overline{L'}_{L4}\tilde{L}_{R4}
+\tilde{M}^{e}_{4} \overline{\tilde{e}}_{L4} e'_{R4}
+H.c.
\label{vecmass}
\ee
and the first three primed masses of each fermion type are zero.
The original charge basis and the heavy mass basis are related by unitary mixing matrices,
\be
Q'_{L}=V_{Q_L}{Q}_{L}, \  u'_{R}=V_{u_R}{u}_{R},
\  d'_{R}=V_{d_R}{d}_{R},
\  L'_{L}=V_{L_L}{L}_{L}, 
\  e'_{R}=V_{e_R}{e}_{R}.
\label{primedbasis}
\ee
The unitary mixing matrix which relates the column vector $Q'_L$ of mass
eigenstates (where the first three components are massless and the fourth
component has a mass $\tilde{M}^{Q}_{4}$) to 
the original fields $Q_L$ may be written as,
\be
V_{Q_L}=V^{Q_L}_{34}V^{Q_L}_{24}V^{Q_L}_{14}, 
\label{VQparam}
\ee
where 
\be
	V^{Q_L}_{34} = \pmatr{1&0&0&0\\0&1&0&0\\0&0&c^{Q_L}_{34}&s^{Q_L}_{34}e^{-i\delta^{Q_L}_{34}}
	\\ 0&0&-s^{Q_L}_{34}e^{i\delta^{Q_L}_{34}}&c^{Q_L}_{34}}, \quad 
\ee
\be
	V^{Q_L}_{24} = \pmatr{1&0&0&0\\0&c^{Q_L}_{24}&0&s^{Q_L}_{24}e^{-i\delta^{Q_L}_{24}}\\0&0&1&0\\
	0&-s^{Q_L}_{24}e^{i\delta^{Q_L}_{24}}&0&c^{Q_L}_{24}}, \quad 
\ee
\be
	V^{Q_L}_{14} = \pmatr{
	c^{Q_L}_{14}&0&0&s^{Q_L}_{14}e^{-i\delta^{Q_L}_{14}}\\ 
	0&1&0&0\\
	0&0&1&0\\
	-s^{Q_L}_{14}e^{i\delta^{Q_L}_{14}}&0&0&c^{Q_L}_{14}}.
	\quad 
\ee
Ignoring phases, the tangent of the mixing angles $t = \tan \theta $ are given by,
\be
t^{Q_L}_{14}=\frac{M^Q_1}{M^Q_4}, \  t^{Q_L}_{24}=\frac{M^Q_2}{M'^Q_4}, \ t^{Q_L}_{34}=\frac{M^Q_1}{M''^Q_4}, 
\ee
where 
\be
M'^Q_4=\sqrt{{M^Q_1}^2+{M^Q_4}^2}, \ M''^Q_4=\sqrt{{M^Q_2}^2+{M'^Q_4}^2}, \ 
\tilde{M}^Q_4=\sqrt{{M^Q_3}^2+{M''^Q_4}^2}.
\ee
Similar equations may be readily obtained in each of the other sectors $u_R,d_R,L_L,e_R$,
with the trivial replacements, $Q_L\rightarrow u_R,d_R,L_L,e_R$.

\subsection{The Lagrangian in the heavy mass basis}
In this subsection we present the Lagrangian of the general class of models in the heavy mass basis, denoted by primes,
in which only the fourth family is heavy (compared to the weak scale).
In this basis the model involves the three massless chiral families $\psi'_i$,
where $i=1,\ldots ,3$ which are massless before electroweak
symmetry breaking, plus a heavy fourth family $\psi'_4$, which have the same chirality as the
first three families, with which they mix.
In this basis, only the fourth family $\psi'_4$ have 
explicit vector-like Dirac mass terms involving the opposite chirality heavy heavy fourth vector-like family $\tilde{\psi}_4$.
The diagonal heavy mass (primed) basis is therefore the correct basis to work in if one wishes to study the interactions of the heavy vector-like 
fourth family states, $\psi'_4$, $\tilde{\psi}_4$, or to integrate them out 
to leave the three massless (before electroweak symmetry breaking) families $\psi'_i$.

\subsubsection{Yukawa couplings and CKM }
In the original basis, the Yukawa couplings in Eq.\ref{yuk} may be written in terms of the three chiral families $\psi_i$ plus the same chirality 
fourth family $\psi_4$ in a $4\times 4$ matrix notation as,
\be
{\cal L}^{Yuk}=H_u\overline{Q}_{L }\tilde{y}^u u_{R }
+H_d\overline{Q}_{L}\tilde{y}^dd_{R}
+H_d\overline{L}_{L}\tilde{y}^ee_{R} +H.c.
\ee
where $\tilde{y}^u,\tilde{y}^d,\tilde{y}^e$ are $4\times 4$ matrices consisting of the original $3\times 3$ matrices, $y^u,y^d,y^e$,
but augmented by a fourth row and column consisting of all zeroes, since we have assumed that the fourth family 
$\psi_4$ does not couple to the Higgs doublets
due to its non-universal $U(1)'$ charges.

In the primed basis in Eq.\ref{primedbasis}, where only the fourth components of the fermions are very heavy,
the Yukawa couplings become,
\be
{\cal L}^{Yuk}=H_u\overline{Q'}_{L }\tilde{y}'^u u'_{R }
+H_d\overline{Q'}_{L } \tilde{y}'^d d'_{R }
+H_d\overline{L'}_{L } \tilde{y}'^e e'_{R }
+H.c.
\ee
where 
\be
\tilde{y}'^u =V_{Q_L}\tilde{y}^u V_{u_R}^{\dagger}, \ \   
\tilde{y}'^d =V_{Q_L}\tilde{y}^d V_{d_R}^{\dagger}, \ \
\tilde{y}'^e =V_{L_L}\tilde{y}^e V_{e_R}^{\dagger}
\label{ytp}
\ee
This shows that the fourth family states $\psi'_4$ with heavy vector-like masses in Eq.\ref{vecmass}
couple to the Higgs by virtue of their mixing with the first three chiral families.

The coupling of the heavy mass eigenstate $\psi'_4$ to the Higgs doublets is given by the fourth rows and columns of the primed
Yukawa matrices in Eq.\ref{ytp},
\bea
{\cal L}^{Yuk}_{heavy}&=&\tilde{y}'^u_{i4}H_u\overline{Q'}_{Li } u'_{R4 }
+ \tilde{y}'^d_{i4}H_d\overline{Q'}_{Li } d'_{R4 }
+\tilde{y}'^e_{i4}H_d\overline{L'}_{Li } e'_{R4 } \nonumber \\
&+&
\tilde{y}'^u_{4i}H_u\overline{Q'}_{L4 } u'_{Ri }
+ \tilde{y}'^d_{4i}H_d\overline{Q'}_{L4 } d'_{Ri }
+\tilde{y}'^e_{4i}H_d\overline{L'}_{L4 } e'_{Ri }
 \nonumber \\
&+&
\tilde{y}'^u_{44}H_u\overline{Q'}_{L4 } u'_{R4}
+ \tilde{y}'^d_{44}H_d\overline{Q'}_{L4 } d'_{R4}
+\tilde{y}'^e_{44}H_d\overline{L'}_{L4 } e'_{R4 }
+H.c.
\label{Yukheavy}
\eea
where 
\bea
&& \tilde{y}'^u_{i4} =  (V_{Q_L}\tilde{y}^u V_{u_R}^{\dagger})_{i4}, \ \   
\tilde{y}'^d_{i4} =(V_{Q_L}\tilde{y}^d V_{d_R}^{\dagger})_{i4}, \ \
\tilde{y}'^e_{i4} =(V_{L_L}\tilde{y}^e V_{e_R}^{\dagger})_{i4}
 \nonumber \\
&& \tilde{y}'^u_{4i} =  (V_{Q_L}\tilde{y}^u V_{u_R}^{\dagger})_{4i}, \ \   
\tilde{y}'^d_{4i} =(V_{Q_L}\tilde{y}^d V_{d_R}^{\dagger})_{4i}, \ \
\tilde{y}'^e_{4i} =(V_{L_L}\tilde{y}^e V_{e_R}^{\dagger})_{4i}
 \nonumber \\
 && \tilde{y}'^u_{44} =  (V_{Q_L}\tilde{y}^u V_{u_R}^{\dagger})_{44}, \ \   
\tilde{y}'^d_{44} =(V_{Q_L}\tilde{y}^d V_{d_R}^{\dagger})_{44}, \ \
\tilde{y}'^e_{44} =(V_{L_L}\tilde{y}^e V_{e_R}^{\dagger})_{44}
\label{ytp4}
\eea
which shows that there will be some Yukawa induced mass mixing between heavy fourth family fermions and light fermions.
This Lagrangian also generates Feynman rules for Higgs bosons which couple the heavy fourth family to the three light chiral families.
However the fourth family is too heavy to be produced in Higgs decays.
There will be a contribution to the Standard Model Higgs production cross-section through gluon-gluon fusion triangle diagrams involving the 
fourth heavy family $\psi'_4$. This is unlike the case of a sequential fourth family, which is excluded by Higgs production 
being too large, due to the large Yukawa couplings of the fourth family to the Higgs boson.
By contrast, in the case of the vector-like fourth family here,
the Yukawa couplings to Higgs doublets in Eq.\ref{ytp4} involving the fourth family will be smaller.
This can be readily understood from Eq.\ref{ytp4}, since
$\tilde{y}^u,\tilde{y}^d,\tilde{y}^e$ have zeroes in the fourth row and column,
and so the couplings like $\tilde{y}'^u_{44},\tilde{y}'^d_{44}$ will involve usual Yukawa couplings and will 
be mixing suppressed.

To calculate the CKM matrix, 
it would not be appropriate to diagonalise the primed
Yukawa matrices in Eq.\ref{ytp} since this would re-mix the heavy vector-like masses throughout all the four families,
and undo the heavy mass diagonalisation. 
The correct proceedure is to integrate out the heavy vector-like family $\psi'_4$, then calculate the CKM matrix in the low
energy effective theory below the heavy vector mass scale.
In the limit of large vector-like masses, ignoring the very small Higgs induced mixing between the heavy fourth family
and the light three families,
one may decouple the heavy states $\psi'_4$,
by simply removing the fourth rows and colums of the primed
Yukawa matrices in Eq.\ref{ytp}, to leave the upper $3\times 3$ blocks, which describe the three massless families,
in the low energy effective theory involving the massless fermions $\psi'_i$,
\be
{\cal L}^{Yuk}_{light}=y'^u_{ij}H_u\overline{Q'}_{Li}u'_{Rj}+y'^d_{ij}H_d\overline{Q'}_{Li}d'_{Rj}
+y'^e_{ij}H_d\overline{L'}_{Li}e'_{Rj} +H.c.
\label{yukp}
\ee
where 
\be
{y}'^u_{ij} =(V_{Q_L}\tilde{y}^u V_{u_R}^{\dagger})_{ij}, \ \   
{y}'^d_{ij} =(V_{Q_L}\tilde{y}^d V_{d_R}^{\dagger})_{ij}, \ \
{y}'^e_{ij} =(V_{L_L}\tilde{y}^e V_{e_R}^{\dagger})_{ij}
\label{yp}
\ee
and $i,j=1,\ldots ,3$. The physical three family quark and lepton masses in the low energy effective theory should be calculated 
using the $3\times 3$ Yukawa matrices in Eq.\ref{yp}.

The CKM matrix for the quarks may be constructed in the usual way, by diagonaling the Yukawa matrices,
$y'^u,y'^d$, 
\be
V'_{uL}y'^uV'^{\dagger}_{uR}= {\rm diag}(y_u,y_c,y_t), \ \ 
V'_{dL}y'^dV'^{\dagger}_{dR}= {\rm diag}(y_d,y_s,y_b)
\label{diag}
\ee
to yield the unitary $3\times 3$ CKM matrix,
\be
V_{\rm CKM}=V'_{uL}V'^{\dagger}_{dL}.
\label{CKM}
\ee
Note that there is no violation of unitarity of the CKM matrix due to the vector-like fourth family.
Also there will be no tree-level Higgs mediated flavour changing neutral currents between the three light families
(the usual GIM mechanism in the Higgs sector).

We emphasise that 
to calculate the CKM matrix and Yukawa eigenvalues one must diagonalise the Yukawa matrices $y'^u,y'^d$ in Eq.\ref{yp},
which emerge after the fourth vector-like family has been correctly decoupled
from the low energy effective theory.
It is incorrect to calculate the CKM matrix from the original Yukawa matrices $y^u,y^d$
in Eq.\ref{yuk}, which do not take into account mixing with the fourth family.

\subsubsection{Gauge couplings}

\underline{Standard Model gauge couplings}

In the diagonal heavy mass (primed) basis, given by the unitary transformations in Eq.\ref{primedbasis},
the gauge Lagrangian in Eq.\ref{gauge} is invariant apart from the $U(1)'$ gauge part.
This is because under the Standard Model gauge group all four families have the same charges,
and so the unitary transformations cancel, as in the usual GIM mechanism.
Thus the part of the Lagrangian involving gluons and electroweak gauge bosons remains flavour diagonal
in the primed basis,
\bea
{\cal L}^{gauge}_{SM}&=&i \overline{Q}'_{L\alpha}\left(\partial_{\mu} - i g_3G^A_{\mu}\frac{\lambda^A}{2}
- i g_2W^a_{\mu}\frac{\sigma^a}{2}-\frac{1}{6}ig_1B_{\mu}\right)\gamma^{\mu}Q'_{L\alpha} \nonumber \\
&+&
i \overline{u}'_{R\alpha}\left(\partial_{\mu} - i g_3G^A_{\mu}\frac{\lambda^A}{2}
 -\frac{2}{3}ig_1B_{\mu}\right)\gamma^{\mu}u'_{R\alpha} \nonumber \\
&+&
i \overline{d}'_{R\alpha}\left(\partial_{\mu} - i g_3G^A_{\mu}\frac{\lambda^A}{2}
 +\frac{1}{3}ig_1B_{\mu}\right)\gamma^{\mu}d'_{R\alpha} \nonumber \\
&+&
i \overline{L}'_{L\alpha}\left(\partial_{\mu} - i g_2W^a_{\mu}\frac{\sigma^a}{2}+\frac{1}{2}ig_1B_{\mu}\right)\gamma^{\mu}L'_{L\alpha} \nonumber \\
&+&
i \overline{e}'_{R\alpha}\left(\partial_{\mu} +ig_1B_{\mu}
\right)\gamma^{\mu}e'_{R\alpha} 
\label{gaugep}
\eea
where $\alpha =1,\ldots ,4$ labels the four families of the same chirality.

The $W^{\pm}$ gauge boson couplings are thus the same as in the Standard Model, 
\be
{\cal L}^{int}_W=\frac{g_2}{\sqrt{2}}\overline{u}'_{L \alpha} W^+_{\mu}\gamma^{\mu}  {d}'_{L\alpha}
+\frac{g_2}{\sqrt{2}}\overline{e}'_{L \alpha} W^+_{\mu}\gamma^{\mu}  \nu'_{L\alpha}
+ H.c.
\ee
where $\alpha =1,\ldots ,4$ labels the four families of the same chirality.
In addition there are the fourth family couplings involving the opposite chirality states $ \tilde{\psi}'_{4}$.
For the quarks, for example, these couplings may be divided into the light three families,
and those involving the fourth family couplings,  
\be
{\cal L}^{int}_W=\frac{g_2}{\sqrt{2}}\overline{u}'_{L i} W^+_{\mu}\gamma^{\mu}  {d}'_{L i}
+\frac{g_2}{\sqrt{2}}\overline{u}'_{L 4} W^+_{\mu}\gamma^{\mu}  {d}'_{L 4}
+\frac{g_2}{\sqrt{2}}\overline{\tilde{u}}'_{R 4} W^+_{\mu}\gamma^{\mu}  {\tilde{d}}'_{R 4}+ H.c.
\ee
and similarly for the leptons. The above couplings allow a fourth family fermion
to decay into $W$ plus a light fermion, after including a small Higgs induced mass insertion 
between a fourth fermion and a light fermion, as shown 
in Eq.\ref{Yukheavy}.

In the low energy effective theory, after the heavy fourth family decouples,
and electroweak symmetry is broken, and the light effective Yukawa matrices are diagonalised
as in Eq.\ref{diag},
the $W$ couplings become,
\be
{\cal L}^{int}_W=\frac{g_2}{\sqrt{2}}\pmatr{\overline{u}_{L} & \overline{c}_{L} & \overline{t}_{L} }V_{\rm CKM}W^+_{\mu}\gamma^{\mu} \pmatr{ d_L  \\ s_L \\ b_L}+ H.c.
\ee
where the CKM matrix is calculated as in Eq.\ref{CKM}.

After electroweak symmetry breaking, the $Z$ gauge boson couples in a flavour diagonal way to all the four
families, both the light families $\psi'_i$ and the heavy family $\psi'_4$, as in the usual GIM mechanism.
This leads to the usual $Z$ interaction Lagrangian
\be
{\cal L}^{int}_Z=\frac{e}{2s_Wc_W}\overline{\psi}'_{\alpha} Z_{\mu}\gamma^{\mu} (C_V^{\psi}-C_A^{\psi}\gamma_5)   {\psi}'_{\alpha}
\label{Zint}
\ee
where 
\be
C_A^{\psi}=t_3, \ \ \ \ C_V^{\psi}=t_3-2s_W^2Q
\ee
where $\psi'_{\alpha} = u'_{\alpha},d'_{\alpha},e'_{\alpha},\nu'_{\alpha}$, where $\alpha =1,\ldots ,4$ labels the four families of the same chirality, and $t_3$ are eigenvalues of $\sigma_3/2$,
while $Q$ are 
the electric charges of the fermions.
We emphasise that this is flavour diagonal, i.e. $Z$ boson exchange does not change the flavour $\alpha$ of a fermion $ {\psi}'_{\alpha}$
in the primed basis.
In particular the heavy fourth family fermions $ {\psi}'_{4}$ thus couple to $Z$ bosons with exactly the same Feynman rules 
as the three light family fermions ${\psi}'_{i}$. 

After the diagonalisation of the light fermion mass matrices, the $Z$ boson couplings remain flavour diagonal,
due to the unitary transformations cancelling, 
and are identical to those in the Standard Model,
namely those in Eq.\ref{Zint}, with the fields $  {\psi}'_{\alpha}$ replaced by their three family mass eigenstates.
The small Higgs induced mass mixing between $ {\psi}'_{4}$ and ${\psi}'_{i}$ will also not lead to any $Z$ induced flavour changing
since any mixing effect will be unitary and will cancel in Eq.\ref{Zint}. We emphasise that such a $Z$ exchange GIM mechanism is a
consequence of the fact that all four families have the same electroweak charges.

\underline{$Z'$ gauge couplings}

The above GIM mechanism in the electroweak sector is in marked contast to the physics of $Z'$ gauge bosons,
where the $U(1)'$ charges depend on the family index $\alpha$. This leads to flavour changing due to 
$Z'$ gauge boson exchange, as we discuss.
After $U(1)'$ breaking, we have a massive $Z'$ gauge boson with diagonal gauge couplings
to the four families of quarks and leptons,
in the original basis,
\be
{\cal L}^{gauge}_{Z'}= g'Z'_{\mu}\left(
\overline{Q}_LD_Q\gamma^{\mu}{Q}_L
+ \overline{u}_RD_u\gamma^{\mu}u_R
+\overline{d}_RD_d\gamma^{\mu}d_R 
+\overline{L}_LD_L\gamma^{\mu}L_L
+\overline{e}_RD_e\gamma^{\mu}e_R
\right)
\label{gaugeZp}
\ee
where 
\bea
&&D_Q={\rm diag}(q_{Q1}, q_{Q2}, q_{Q3},q_{Q4}), \ 
D_u={\rm diag}(q_{u1}, q_{u2}, q_{u3},q_{u4}), \ 
D_d={\rm diag}(q_{d1}, q_{d2}, q_{d3},q_{d4}), \nonumber \\
&&D_L={\rm diag}(q_{L1}, q_{L2}, q_{L3},q_{L4}), \ 
D_e={\rm diag}(q_{e1}, q_{e2}, q_{e3},q_{e4}).
\label{Zpcharges}
\eea
In the diagonal heavy mass (primed) basis, given by the unitary transformations in Eq.\ref{primedbasis},
the $Z'$ couplings to the four families of quarks and leptons in Eq.\ref{gaugeZp} becomes,
\be
{\cal L}^{gauge}_{Z'}= g'Z'_{\mu}\left(
\overline{Q}'_LD'_Q\gamma^{\mu}{Q}'_L
+ \overline{u}'_RD'_u\gamma^{\mu}u'_R
+\overline{d}'_RD'_d\gamma^{\mu}d'_R 
+\overline{L}'_LD'_L\gamma^{\mu}L'_L
+\overline{e}'_RD'_e\gamma^{\mu}e'_R
\right)
\label{gaugeZp1}
\ee
where 
\bea
&&D'_Q= V_{Q_L}D_QV_{Q_L}^{\dagger}, \ 
D'_u=V_{u_R}D_uV_{u_R}^{\dagger}, \ 
D'_d=V_{d_R}D_dV_{d_R}^{\dagger}, \nonumber \\
&&D'_L=V_{L_L}D_LV_{L_L}^{\dagger}, \ 
D'_e=V_{e_R}D_eV_{e_R}^{\dagger}.
\eea
Although the $4\times 4$ matrices $D_Q$, etc., are diagonal in flavour space, the 
$4\times 4$ matrices $D'_Q$, etc., are not generally diagonal in flavour space,
since the $U(1)'$ charges may be different for the four flavours.
This is the case even if the $U(1)'$ charges are universal for the first three families, but differ only for the fourth family.
Recall that in the primed basis the fourth family is very heavy while the first three are light.
Then Eq.\ref{gaugeZp1} shows that, in general, $Z'$ exchange can couple two light families of different flavour,
or a heavy fourth family fermion to a light fermion of the first three families.
For example, a $Z'$ exchange diagram will allow the decay of a heavy fourth family fermion to three light fermions of 
different flavours. This decay mechanism will compete with the decay of a heavy fourth family fermion into 
a $W$ plus a light fermion, which is suppressed by the small Higgs induced mass insertion arising from Eq.\ref{Yukheavy}.
 
In the low energy effective theory, after decoupling the fourth heavy family, Eq.\ref{gaugeZp1}
gives the $Z'$ couplings to the three massless families of quarks and leptons,
\be
{\cal L}^{gauge}_{Z'}= g'Z'_{\mu}\left(
\overline{Q}'_{L}\tilde{D}'_Q\gamma^{\mu}{Q}'_{L}
+ \overline{u}'_{R}\tilde{D}'_u\gamma^{\mu}u'_{R}
+\overline{d}'_{R}\tilde{D}'_d\gamma^{\mu}d'_{R}
+ \overline{L}'_{L}\tilde{D}'_L\gamma^{\mu}L'_{L}
+\overline{e}'_{R}\tilde{D}'_e\gamma^{\mu}e'_{R}
\right)
\label{gaugeZp2}
\ee
where the $3\times 3$ matrices $\tilde{D}'$ are given by,
\bea
&&(\tilde{D}'_Q)_{ij}= (V_{Q_L}D_QV_{Q_L}^{\dagger})_{ij}, \ 
(\tilde{D}'_u)_{ij}=(V_{u_R}D_uV_{u_R}^{\dagger})_{ij}, \ 
(\tilde{D}'_d)_{ij}=(V_{d_R}D_dV_{d_R}^{\dagger})_{ij}, \nonumber \\
&&(\tilde{D}'_L)_{ij}=(V_{L_L}D_LV_{L_L}^{\dagger})_{ij}, \ 
(\tilde{D}'_e)_{ij}=(V_{e_R}D_eV_{e_R}^{\dagger})_{ij},
\label{Dp}
\eea
where $i,j=1,\ldots ,3$.
We emphasise that these matrices are not diagonal,
leading to flavour changing neutral currents, mediated by tree-level $Z'$ exchange.
In the parametrisation in Eq.\ref{VQparam}, ignoring phases, each of the symmetric 
$3\times 3$ matrices $\tilde{D}'$ schematically looks like,
\be
\footnotesize{
	\tilde{D}'= \pmatr{q_{1}c^2_{14}+q_{4}s^2_{14}&s_{14}s_{24}c_{14}(q_4-q_1)&(s_{14}s_{34}c_{14}c_{24})(q_4-q_1)\\
	.&q_{1}s^2_{14}s^2_{24}+q_{2}c^2_{24}+q_{4}s^2_{24}c^2_{14}
	&q_{1}s^2_{14}s_{24}s_{34}c_{24}-q_{2}s_{24}s_{34}c_{24}+q_{4}s_{24}s_{34}c_{24}c^2_{14}\\
	.&.&q_{1}s^2_{14}s^2_{34}c^2_{24}+q_{2}s^2_{24}s^2_{34}+q_{3}c^2_{34}+q_{4}s^2_{34}c^2_{14}c^2_{24}}
	}
\label{Dpexplicit}
\ee
with different angles and charges for each matrix in Eq.\ref{Dp}.
When $q_1=q_2=q_3=q_4$ these matrices are proportional to the unit matrix and there is no flavour changing
due to $Z'$ exchange.
Also when $s_{i4}=\sin \theta_{i4}=0$, these matrices are flavour diagonal.

After diagonalisation of the light quark Yukawa matrices, as in Eq.\ref{diag},
the $Z'$ couplings to the physical quark mass eigenstates $u,c,t,d,s,b$ are given from Eq.\ref{gaugeZp2} by,
\bea
{\cal L}^{q}_{Z'}&=& 
g'Z'_{\mu}\pmatr{\overline{u}_{L} & \overline{c}_{L} & \overline{t}_{L} } V'_{uL} \tilde{D}'_Q V'^{\dagger}_{uL}\gamma^{\mu} \pmatr{u_L  \\ c_L \\ t_L}
\nonumber \\
&+& 
g'Z'_{\mu}\pmatr{\overline{d}_{L} & \overline{s}_{L} & \overline{b}_{L} } V'_{dL} \tilde{D}'_Q V'^{\dagger}_{dL}\gamma^{\mu} \pmatr{d_L  \\ s_L \\ b_L}
\nonumber \\
&+& 
g'Z'_{\mu}\pmatr{\overline{u}_{R} & \overline{c}_{R} & \overline{t}_{R} } V'_{uR} \tilde{D}'_u V'^{\dagger}_{uR}\gamma^{\mu} \pmatr{u_R  \\ c_R \\ t_R}
\nonumber \\
&+& 
g'Z'_{\mu}\pmatr{\overline{d}_{R} & \overline{s}_{R} & \overline{b}_{R} } V'_{dR} \tilde{D}'_d V'^{\dagger}_{dR}\gamma^{\mu} \pmatr{d_R  \\ s_R \\ b_R}
\label{gaugeZp3}
\eea
Similarly the charged lepton couplings to $Z'$ will be given by analogous results,
\bea
{\cal L}^{e}_{Z'}&=& 
g'Z'_{\mu}\pmatr{\overline{e}_{L} & \overline{\mu}_{L} & \overline{\tau}_{L} } V'_{eL} \tilde{D}'_L V'^{\dagger}_{eL}\gamma^{\mu} 
\pmatr{e_L  \\ \mu_L \\ \tau_L}
\nonumber \\
&+& 
g'Z'_{\mu}\pmatr{\overline{e}_{R} & \overline{\mu}_{R} & \overline{\tau}_{R} } V'_{eR} \tilde{D}'_e V'^{\dagger}_{eR}\gamma^{\mu} 
\pmatr{e_R  \\ \mu_R \\ \tau_R}
\label{gaugeZp4}
\eea
Finally, ignoring neutrino mass, the $Z'$ couplings to left-handed neutrinos are given by,
\be
{\cal L}^{\nu}_{Z'}= 
g'Z'_{\mu}\pmatr{\overline{\nu_e}_{L} & \overline{\nu_\mu}_{L} & \overline{\nu_\tau}_{L} }   V'_{eL} \tilde{D}'_L V'^{\dagger}_{eL} \gamma^{\mu} 
\pmatr{\nu_{eL}  \\ \nu_{\mu L} \\ \nu_{\tau L}}
\label{gaugeZp45}
\ee
These results show that, if the $\tilde{D}'$ term is proportional to the unit matrix,
then this will not lead to flavour violation. 
However any non-universal part of $\tilde{D}'$
will lead to flavour changing
in the physical mass basis of the light fermions.
We shall see explicit examples of the application of this formalism in the next section.

\section{Examples of flavourful $Z'$ Models }
\label{examples}
The results in the previous section are of quite general applicability.
However, to illustrate the mechanism and show how the formalism may be applied in practice,
it is instructive to consider two concrete examples of well known $Z'$ models which can be made flavourful
via mixing with a non-universal fourth vector-like family and show how they can
provide an explanation of $R_K$ and $R_{K^*}$. Clearly the same method could be applied to 
any $Z'$ model including $B-L$ models, $E_6$ models,
composite models, and so on.

\subsection{Fermiophobic model}

The first example we consider is one in which the quarks and leptons start out not coupling to the $Z'$ at all,
as in fermiophobic models. We show that such fermiophobic $Z'$ models  
may be converted to flavourful $Z'$ models via mixing with a fourth vector-like family with $Z'$ couplings.
We then show how such a model is capable of accounting for $R_K$ and $R_{K^*}$.

The starting point is a class of fermiophobic models, where
none of the three chiral families of quarks and leptons (nor the Higgs doublets) carry the $U(1)'$
charges, together with a fourth vector-like family which carry $U(1)'$ charges, i.e. $q_1=q_2=q_3=0$ but $q_4\neq 0$.
The charges in Table~\ref{tab:funfields} are therefore given by the diagonal matrices in Eq.\ref{Zpcharges}:
\bea
&&D_Q={\rm diag}(0, 0, 0,q_{Q4}), \ 
D_u={\rm diag}(0, 0, 0,q_{u4}), \ 
D_d={\rm diag}(0,0,0,q_{d4}), \nonumber \\
&&D_L={\rm diag}(0,0,0,q_{L4}), \ 
D_e={\rm diag}(0,0,0,q_{e4}).
\label{Zpcharges1}
\eea
In addition we assume Higgs singlets $\phi_{\psi}$ 
with charges $|q_{\phi_\psi}| =|q_{\psi_4} |$ whose VEVs yield a massive $Z'$,
and whose couplings permit mixing of the fourth vector-like family with the three families of the same chirality.
The mixing of quarks and leptons with the fourth vector-like family induces 
flavour violating $Z'$ couplings to the three light families of quarks and leptons, as in Eq.\ref{gaugeZp2}, which depend on
$3\times 3$ matrices $\tilde{D}'$ in Eq.\ref{Dpexplicit} of the form,
\be
\tilde{D}'_Q= q_{Q4}\pmatr{(s^{Q}_{14})^2&s^{Q}_{14}s^{Q}_{24}c^{Q}_{14}&s^{Q}_{14}s^{Q}_{34}c^{Q}_{14}c^{Q}_{24}\\
	.&(s^{Q}_{24})^2(c^{Q}_{14})^2
	&s^{Q}_{24}s^{Q}_{34}c^{Q}_{24}(c^{Q}_{14})^2\\
	.&.&(s^{Q}_{34})^2(c^{Q}_{14})^2(c^{Q}_{24})^2},  
	\label{Dpexplicit2}
\ee
and similar matrices with $Q\rightarrow L$, and so on.
The couplings of the quark and lepton mass eigenstates to the $Z'$ are given by inserting Eq.\ref{Dpexplicit2},
and similar equations in each of the sectors $Q_L,u_R,d_R,L_L,e_R$, into
Eqs.~\ref{gaugeZp3}, \ref{gaugeZp4}, \ref{gaugeZp45}.
This shows that the $Z'$ will couple in a flavour violating way to the three light families, even though they carry no $U(1)'$ charges,
because of their mixing with the fourth family which do carry $U(1)'$ charges. The mixing is controlled by three mixing angles
$\theta_{i4}$ in each of the sectors $Q_L,u_R,d_R,L_L,e_R$, which involves 15 parameters.

Assuming that only $\theta^{Q_L}_{34}$ and $\theta^{L_L}_{34}$ are non-zero,
with all other mixing angles being zero, the mixing matrices in Eq.\ref{Dpexplicit2} become,
\be
	\tilde{D}'_Q=q_{Q4} \pmatr{0&0&0\\
	0&0&0\\
	0&0&(s^{Q}_{34})^2},  \ \ 
	\tilde{D}'_L= q_{L4} \pmatr{0&0&0\\
	0&0&0\\
	0&0&(s^{L}_{34})^2}
	\label{Dpexplicit3}
\ee
so that the $Z'$ couplings from Eq.\ref{gaugeZp2} become,
\be
{\cal L}^{gauge}_{Z'}= g'Z'_{\lambda}\left(
q_{Q4}(s^{Q}_{34})^2 \overline{Q}'_{L3}\gamma^{\lambda}{Q}'_{L3}
+q_{L4}(s^{L}_{34})^2 \overline{L}'_{L3}\gamma^{\lambda}L'_{L3}
\right)
\label{gaugeZp7}
\ee
where the $Z'$ couples to the third family left-handed quark and lepton doublets ${Q}'_{L3}=(t'_L,b'_L)$ 
and ${L}'_{L3}=(\nu'_{\tau L},\tau'_L)$, where the primes indicate that these are the states before the 
Yukawa matrices are diagonalised.
In particular this will lead the couplings,
\bea
{\cal L}^{gauge}_{Z'}&= &g'Z'_{\lambda}\left(
q_{Q4}(s^{Q}_{34})^2 \overline{b}'_{L}\gamma^{\lambda}{b}'_{L}
+ q_{L4}(s^{L}_{34})^2 \overline{\tau}'_{L}\gamma^{\lambda}\tau'_{L}+\ldots
\right),\nonumber  \\
&\approx &g'Z'_{\lambda}\left(
q_{Q4}(s^{Q}_{34})^2(V'^{\dagger}_{dL})_{32}  \overline{b}_{L}\gamma^{\lambda}{s}_{L}
+ q_{L4}(s^{L}_{34})^2 |(V'^{\dagger}_{eL})_{32} |^2 \overline{\mu}_{L}\gamma^{\lambda}\mu_{L}+\ldots
\right),
\label{gaugeZp8}
\eea
where we have used Eq.~\ref{diag} to expand the primed fields in terms of mass eigenstates,
\bea
b'_L&=&(V'^{\dagger}_{dL})_{31} d_L + (V'^{\dagger}_{dL})_{32} s_L + (V'^{\dagger}_{dL})_{33} b_L , \nonumber \\
\tau'_L&=&(V'^{\dagger}_{eL})_{31} e_L + (V'^{\dagger}_{eL})_{32} \mu_L + (V'^{\dagger}_{eL})_{33} \tau_L ,
\eea
and assumed from 
the hierarchy of the CKM matrix that, 
\bea
|(V'^{\dagger}_{dL})_{31}|^2 \ll
|(V'^{\dagger}_{dL})_{32} |^2 \ll (V'^{\dagger}_{dL})_{33}^2 &\approx& 1, \nonumber \\
|(V'^{\dagger}_{eL})_{31}|^2 \ll
|(V'^{\dagger}_{eL})_{32} |^2 \ll (V'^{\dagger}_{eL})_{33}^2 &\approx& 1.
\label{hierarchy1}
\eea
From Eq.~\ref{gaugeZp8}, $Z'$ exchange generates the effective operator, as in Eq.~\ref{G},
\be\label{eq:HNP} 
G^{\rm BSM}_{b_L\mu_L}\, \bar b_L\gamma^\lambda s_L \, \bar \mu_L \gamma_\lambda \mu_L\,, 
\ee
where we identify,
\be
G^{\rm BSM}_{b_L\mu_L} \; =q_{Q4}q_{L4}(s^{Q}_{34})^2(s^{L}_{34})^2(V'^{\dagger}_{dL})_{32} |(V'^{\dagger}_{eL})_{32} |^2 \left(\frac{g'^2}{{M_Z'}^2}\right) .
\label{G1}
\ee
This operator dominates over the analogous operator with $\mu_L$ replaced by $e_L$,
according to Eq.\ref{hierarchy1}. To explain the $R_K$ and $R_{K^*}$ anomalies we require $G^{\rm BSM}_{b_L\mu_L}$ to have the correct sign
and magnitude, as discussed in Eqs.~\ref{G}, \ref{33}.
Motivated by the CKM matrix, we may assume $(V'^{\dagger}_{dL})_{32}$ and $(V'^{\dagger}_{eL})_{32}$ are both of order 
 $V_{ts} \approx 0.04$. Then Eq.~\ref{33} requires,
 \be
G^{\rm BSM}_{b_L\mu_L} \; = 
q_{Q4}q_{L4}(s^{Q}_{34})^2(s^{L}_{34})^2 (6\times 10^{-5}) \left(\frac{g'^2}{{M_{Z'}}^2}\right)
 \approx -\frac{1}{(33 \ {\rm TeV})^2}
\label{331}
\ee
This suggests that in this model $M_{Z'}\lesssim (\sqrt{10^{-5}})\times  33$ TeV $\lesssim 100$ GeV.
Such a light $Z'$ is not excluded since it does not couple to first generation quarks and leptons, so would not be produced
at LEP, and its Drell-Yan production at the LHC would only proceed via $\bar s s$ annihilation through a coupling which is 
amplitude suppressed by
$(V'^{\dagger}_{dL})^2_{32}\sim V_{ts}^2 \sim10^{-3}$.

\subsection{$SO(10)$ model }

The next example we consider is an $SO(10)$ model which breaks at the GUT scale,
\be
SO(10)\rightarrow SU(5)\times U(1)_{\chi}
\ee
under which the three chiral $16$ representations decompose as,
\be
{\bf 16}_i\rightarrow ({\bf 10}, 1)_{i}+({\bf \overline{5}}, -3)_i+({\bf 1},5)_{i}
\ee
where $U(1)_{\chi}$ charges should all be multipled by a normalisation factor of $\frac{1}{2\sqrt{10}}$.
The $U(1)_{\chi}$ survives to low energy and is broken at the few TeV scale to provide an observable $Z'$.
We assume that the low energy fourth vector-like family arises from incomplete surviving parts of the decompositions,
\bea
{\bf 45} &\rightarrow &({\bf 24}, 0)+ ({\bf 10}, -4)+({\bf \overline{10}}, 4)+({\bf 1},0)\nonumber \\
 {\bf 10}&\rightarrow &({\bf 5}, -2)+({\bf \overline{5}}, 2).
\eea
The fourth vector-like family with masses near the few TeV scale consists of the following surviving parts of these multiplets,
\be
({\bf 10}, -4)+({\bf \overline{10}}, 4)+({\bf 5}, -2)+({\bf \overline{5}}, 2),
\label{vectorfamily}
\ee
where we assume that the other $({\bf 24}, 0)$ and $({\bf 1},0)$ parts get large GUT scale masses.
The three chiral families and the fourth vector-like family are odd under a matter parity.
The Higgs doublets emerge from a different ${\bf 10}_H$ with even matter parity, 
allowing Higgs Yukawa couplings. In addition we will need the Higgs ${\bf 16}_H$ and $\overline {\bf 16}_H$
with even matter parity to mix the fourth vector-like family with the three chiral families. 
We do not address any doublet-triplet or other splitting problems here.

Then $SU(5)$ subsequently breaks to the Standard Model gauge group at the GUT scale,
\be
{\bf 10}\rightarrow Q,u^c,e^c, \ \ \ \ {\bf \overline{5}} \rightarrow L,d^c
\ee
We emphasise that the single vector-like family in Eq.\ref{vectorfamily},
includes quark and lepton doublets necessary to account for $R_K$ and $R_{K^*}$.
\footnote{This may be compared the $SO(10)$ model in \cite{Hisano:2015pma}
where there are three low energy $({\bf 5}, -2)+({\bf \overline{5}}, 2)$ representations mixing with the three chiral families
leading to flavour changing $Z'$ interactions. However such a model 
is unable to account for $R_K$ and $R_{K^*}$, in the absence of vector-like quark doublets.}
In terms of the fields $Q_L,u_R,d_R,L_L,e_R$,
the charges under $U(1)'=U(1)_{\chi}$ in Table~\ref{tab:funfields} are therefore given by the diagonal matrices:
\bea
&&D_Q={\rm diag}\left(1, 1, 1, -4\right), \ 
D_u={\rm diag}\left (-1, -1, -1, 4  \right), \ 
D_d={\rm diag}\left( 3, 3, 3, -2 \right), \nonumber \\
&&D_L={\rm diag}\left( -3, -3, -3, 2 \right), \ 
D_e={\rm diag}\left( -1, -1, -1, 4  \right),
\label{Zpcharges2}
\eea
up to a normalisation factor of $\frac{1}{2\sqrt{10}}$ multiplying each matrix.

In addition we assume Higgs singlets 
whose VEVs yield a massive $Z'$,
and whose couplings permit mixing of the fourth vector-like family with the three families of the same chirality.
The mixing of quarks and leptons with the fourth vector-like family induces 
flavour violating $Z'$ couplings to the three light families of quarks and leptons, as in Eq.\ref{gaugeZp2}, which depend on
$3\times 3$ matrices $\tilde{D}'$ from Eq.\ref{Dpexplicit} of the form,
\bea
	\tilde{D}'_L&=& -\frac{3}{2\sqrt{10}}
	\pmatr{1&0&0\\
	0&1&0\\
	0&0&1}
	+ \frac{5}{2\sqrt{10}}\pmatr{(s^{L}_{14})^2&s^{L}_{14}s^{L}_{24}c^{L}_{14}&s^{L}_{14}s^{L}_{34}c^{L}_{14}c^{L}_{24}\\
	.&(s^{L}_{24})^2(c^{L}_{14})^2
	&s^{L}_{24}s^{L}_{34}c^{L}_{24}(c^{L}_{14})^2\\
	.&.&(s^{L}_{34})^2(c^{L}_{14})^2(c^{L}_{24})^2} \nonumber \\
		\tilde{D}'_Q&=& \frac{1}{2\sqrt{10}}
	\pmatr{1&0&0\\
	0&1&0\\
	0&0&1}
	- \frac{5}{2\sqrt{10}}\pmatr{(s^{Q}_{14})^2&s^{Q}_{14}s^{Q}_{24}c^{Q}_{14}&s^{Q}_{14}s^{Q}_{34}c^{Q}_{14}c^{Q}_{24}\\
	.&(s^{Q}_{24})^2(c^{Q}_{14})^2
	&s^{Q}_{24}s^{Q}_{34}c^{Q}_{24}(c^{Q}_{14})^2\\
	.&.&(s^{Q}_{34})^2(c^{Q}_{14})^2(c^{Q}_{24})^2} 
\label{Dpexplicit4}
\eea
Eq.\ref{Dpexplicit4} 
consists of a universal matrix, proportional to the unit matrix, plus a non-universal matrix of the same form as Eq.\ref{Dpexplicit2},
but of opposite sign to that 
which appeared in the fermiophobic model.
Similar matrices may be written down for each of the sectors $Q_L,u_R,d_R,L_L,e_R$.
The couplings of the quark and lepton mass eigenstates to the $Z'$ are given by inserting Eq.\ref{Dpexplicit4},
and similar equations in each of the sectors $Q_L,u_R,d_R,L_L,e_R$, into
Eqs.~\ref{gaugeZp3}, \ref{gaugeZp4}.

Assuming that only $\theta^{Q_L}_{34}$ and $\theta^{L_L}_{14}$ are non-zero,
with all other mixing angles being zero, the mixing matrices in Eq.\ref{Dpexplicit4} simplify,
\bea
	\tilde{D}'_L&=& -\frac{3}{2\sqrt{10}}
	\pmatr{1&0&0\\
	0&1&0\\
	0&0&1}
	+ \frac{5}{2\sqrt{10}}\pmatr{(s^{L}_{14})^2&0&0\\
	0&0&0\\
	0&0&0 } \nonumber \\
		\tilde{D}'_Q&=& \frac{1}{2\sqrt{10}}
	\pmatr{1&0&0\\
	0&1&0\\
	0&0&1}
	- \frac{5}{2\sqrt{10}}\pmatr{0&0&0\\
	0&0&0\\
	0&0&(s^{Q}_{34})^2 }
\label{Dpexplicit5}
\eea
The other matrices are universal, since we assume their mixing angles are zero,
\be
	\tilde{D}'_e=\tilde{D}'_u=-\frac{1}{2\sqrt{10}}
	\pmatr{1&0&0\\
	0&1&0\\
	0&0&1 },\ \ \ \ 
	\tilde{D}'_d=\frac{3}{2\sqrt{10}}
	\pmatr{1&0&0\\
	0&1&0\\
	0&0&1 }
	\label{Dpexplicit6}
\ee
The universal (unit matrix) parts of Eq.\ref{Dpexplicit5} and \ref{Dpexplicit6} when inserted into Eqs.\ref{gaugeZp3}, \ref{gaugeZp4}, \ref{gaugeZp45},
lead to the universal $Z'$ couplings for the quarks,
\bea
{\cal L}^{q, univ}_{Z'}&=& 
\frac{1}{2\sqrt{10}}g'Z'_{\mu}\pmatr{\overline{u}_{L} & \overline{c}_{L} & \overline{t}_{L} } \gamma^{\mu} \pmatr{u_L  \\ c_L \\ t_L}
\nonumber \\
&+& 
\frac{1}{2\sqrt{10}}g'Z'_{\mu}\pmatr{\overline{d}_{L} & \overline{s}_{L} & \overline{b}_{L} } \gamma^{\mu} \pmatr{d_L  \\ s_L \\ b_L}
\nonumber \\
&-& 
\frac{1}{2\sqrt{10}}g'Z'_{\mu}\pmatr{\overline{u}_{R} & \overline{c}_{R} & \overline{t}_{R} } \gamma^{\mu} \pmatr{u_R  \\ c_R \\ t_R}
\nonumber \\
&+& 
\frac{3}{2\sqrt{10}}
g'Z'_{\mu}\pmatr{\overline{d}_{R} & \overline{s}_{R} & \overline{b}_{R} } \gamma^{\mu} \pmatr{d_R  \\ s_R \\ b_R}
\label{gaugeZp11}
\eea
Similarly the charged lepton couplings to $Z'$ will be given by analogous results,
\bea
{\cal L}^{e, univ}_{Z'}&=& 
-\frac{3}{2\sqrt{10}}
g'Z'_{\mu}\pmatr{\overline{e}_{L} & \overline{\mu}_{L} & \overline{\tau}_{L} }\gamma^{\mu} 
\pmatr{e_L  \\ \mu_L \\ \tau_L}
\nonumber \\
&-& 
\frac{1}{2\sqrt{10}}g'Z'_{\mu}\pmatr{\overline{e}_{R} & \overline{\mu}_{R} & \overline{\tau}_{R} } \gamma^{\mu} 
\pmatr{e_R  \\ \mu_R \\ \tau_R}
\label{gaugeZp12}
\eea
Finally, ignoring neutrino mass, the $Z'$ couplings to left-handed neutrinos are given by,
\be
{\cal L}^{\nu }_{Z'}= 
-\frac{3}{2\sqrt{10}}
g'Z'_{\mu}\pmatr{\overline{\nu_e}_{L} & \overline{\nu_\mu}_{L} & \overline{\nu_\tau}_{L} }  \gamma^{\mu} 
\pmatr{\nu_{eL}  \\ \nu_{\mu L} \\ \nu_{\tau L}}
\label{gaugeZp13}
\ee

There will be also be additional quark and lepton couplings from the non-universal parts of Eq.\ref{Dpexplicit5},
which, when inserted into Eq.\ref{gaugeZp2}, leads to,
\bea
{\cal L}^{nonuniv}_{Z'}&= &
\frac{5}{2\sqrt{10}}g'Z'_{\lambda}\left(  (s^{L}_{14})^2\bar e'_L\gamma^\lambda e'_L -(s^{Q}_{34})^2\bar b'_L\gamma^\lambda b'_L 
+\ldots
\right),  \label{gaugeZp13} \\
&\approx &
\frac{5}{2\sqrt{10}} g'Z'_{\lambda}\left((s^{L}_{14})^2\bar e_L\gamma^\lambda e_L
 -(s^{Q}_{34})^2\bar b_L\gamma^\lambda b_L 
-(s^{Q}_{34})^2 (V'^{\dagger}_{dL})_{32} \bar b_L\gamma^\lambda s_L 
+\ldots
\right),  \nonumber 
\eea
where we have used Eq.~\ref{diag} to expand the primed fields in terms of mass eigenstates,
\bea
b'_L&=&(V'^{\dagger}_{dL})_{31} d_L + (V'^{\dagger}_{dL})_{32} s_L + (V'^{\dagger}_{dL})_{33} b_L \nonumber \\
e'_L&=&(V'^{\dagger}_{eL})_{11} e_L + (V'^{\dagger}_{eL})_{12} \mu_L + (V'^{\dagger}_{eL})_{13} \tau_L 
\eea 
and assumed from 
the hierarchy of the CKM matrix that 
\bea
|(V'^{\dagger}_{dL})_{31}|^2 \ll
|(V'^{\dagger}_{dL})_{32} |^2 \ll (V'^{\dagger}_{dL})_{33}^2 &\approx& 1, \nonumber \\
|(V'^{\dagger}_{eL})_{13}|^2 \ll
|(V'^{\dagger}_{eL})_{12} |^2 \ll (V'^{\dagger}_{eL})_{11}^2 &\approx& 1.
\label{hierarchy2}
\eea

Combining the universal $Z'$ couplings in Eq.~\ref{gaugeZp12} with the non-universal couplings in Eq.~\ref{gaugeZp13},
leads to $Z'$ mediated operators relevant for rare $B$ decays, 
\be
\label{eq:HNP2} 
G^{\rm BSM}_{b_L\mu_L} \ 
\bar b_L\gamma^\lambda s_L \left[ \ 
 \bar\mu_L \gamma_\lambda \mu_L 
 +
\left(1-\frac{5}{3}(s^{L}_{14})^2\right) \bar e_L \gamma_\lambda e_L 
+ \frac{1}{3} \bar\mu_R \gamma_\lambda \mu_R
+ \frac{1}{3} \bar e_R \gamma_\lambda e_R
+\cdots
\right]
\ee
where
\be
G^{\rm BSM}_{b_L\mu_L} = \frac{3}{8}(s^{Q}_{34})^2(V'^{\dagger}_{dL})_{32} \left(\frac{g'^2}{{M_Z'}^2}\right).
\label{G2}
\ee
If $(s^{L}_{14})^2\approx 3/5$ then the $\bar e_L  {e}_L$ couplings will be suppressed.
Also note that the $\bar e_R  {e}_R$ and $\bar\mu_R \mu_R$ couplings are 
$1/3$ times those of $\bar\mu_L \mu_L $, as predicted by $SO(10)$.
Since the $\bar\mu_L \mu_L $ term dominates,
then the model can
explain the $R_K$ and $R_{K^*}$ anomalies, if 
$G^{\rm BSM}_{b_L\mu_L} $ has the correct sign and magnitude,
as in Eqs.~\ref{G}, \ref{33}.
Assuming that 
$g'\approx 0.46$ \cite{Accomando:2010fz}, Eq.\ref{G2} and Eq.\ref{G} then imply,
\be
M_Z' \approx (s^{Q}_{34})\, (V'^{\dagger}_{dL})^{1/2}_{32}\; (9 \; {\rm TeV})
\label{G3}
\ee
Since the $Z'$ in this model has flavour diagonal couplings to muons similar to the usual $U(1)_{\chi}$ model,
the usual LHC limits apply, so we must have $M_{Z'}\gtrsim 3$ TeV \cite{Aaboud:2016cth},
which implies $(s^{Q}_{34})\, (V'^{\dagger}_{dL})^{1/2}_{32}\gtrsim1/3$.
Actually $(s^{Q}_{34})\, (V'^{\dagger}_{dL})^{1/2}_{32}\gtrsim1/3$ is quite a stringent limit,
for example the usual CKM inspired expectation $(V'^{\dagger}_{dL})^{1/2}_{32}\sim \lambda \sim 0.22$
is already not viable, in agreement with the general results in \cite{Greljo:2017vvb}.
However large mixings such as, for example,
$(s^{Q}_{34})\sim 1/\sqrt{2}$ and $(V'^{\dagger}_{dL})^{1/2}_{32}\sim 0.5$,
would imply $M_Z' \sim 3.2 $ TeV, just above the current limit.
Note that the couplings of the $Z'$ to electrons will be suppressed in this model relative to muons,
which is the main LHC prediction of the model. Therefore the model predicts an imminent LHC discovery of 
a $Z'$ in the muon channel, with a suppressed coupling in the electron channel.
In addition, the model predicts $\bar\mu_L e_L$ and $\bar e_L  {\mu}_L$
lepton flavour violating final states, with an amplitude suppressed by $(V'^{\dagger}_{eL})_{21}$,
which is typically of order of a third of the Cabibbo angle in unified models.

\section{Conclusion}
\label{conclusion}

In this paper we have shown how any flavour conserving $Z'$ model
can be made flavour violating and non-universal 
by introducing mass mixing of quarks and leptons with a fourth family of vector-like fermions with non-universal $Z'$ couplings.
We have developed a general formalism to achieve this for any $Z'$ model, including $B-L$ models, $E_6$ models,
composite models, and so on. All that is required is to specify the charges for the model in Table~\ref{1}.
These charges may be conveniently summarised in terms of the charge matrices defined in Eq.\ref{Zpcharges}.
Once these charge matrices are written down for a particular model, 
the Lagrangian is completely specified using the general results given in the paper.

To illustrate the proceedure, we have considered two concrete examples, namely a fermiophobic model, and an 
$SO(10)$ GUT model, and shown how they can account for 
the anomalous $B$ decay ratios
$R_K$ and $R_{K^*}$. 
In both examples, we have simply written down the charge matrices for the models, then applied the general results
of the paper to calculate the Feynman rules for the $Z'$ couplings to physical quark and lepton mass eigenstates,
and isolated the flavour diagonal and off-diagonal parts resonsible for flavour violation and non-universality.
The SM gauge couplings do not violate flavour in this models, since the three chiral quark and lepton families mix with 
the vector-like family with the same SM quantum numbers, only differing due to the non-universality of the $U(1)'$ charges.

The experimental predictions of such models are very rich and varied, and deserve a dedicated 
phenomenological study, beyond the scope of the present paper. Generally speaking, the phenomenological predictions may be divided
into low energy flavour changing and rare processes, and high energy collider signatures.
The low energy flavour changing will encompass lepton flavour violation, including $\tau$ decays \cite{Foldenauer:2016rpi}, 
while the LHC predictions include a $Z'$ as well as a complete fourth vector-like family, with interesting flavour dependent signatures.
The $Z'$ may be light and weakly coupled, for example around $100$ GeV in the fermiophobic model, or heavier with non-universal
couplings to electrons and muons, for example just above the current LHC limit of $3$ TeV in the $SO(10)$ model.

In conclusion, we have proposed a new class of flavourful $Z'$ models which may be obtained as a bolt-on or upgrade to any existing 
anomaly free $Z'$
model, by adding a vector-like fourth family, with non-universal $U(1)'$ charges,
together with scalar singlets which allow mass mixing to take place between the
three chiral families and the vector-like family. We have shown that the resulting low energy $Z'$ couplings will always violate flavour
and may account for the anomalous $B$ decay ratios $R_K$ and $R_{K^*}$.

\subsection*{Acknowledgements}

S.\,F.\,K. acknowledges the STFC Consolidated Grant ST/L000296/1 and the European Union's Horizon 2020 Research and Innovation programme under Marie Sk\l{}odowska-Curie grant agreements Elusives ITN No.\ 674896 and InvisiblesPlus RISE No.\ 690575.

\end{document}